# Questions for a Materialist Philosophy Implying the Equivalence of Computers and Human Cognition

Douglas M. Snyder
Los Angeles, California

ABSTRACT

Issues related to a materialist philosophy are explored as concerns the implied equivalence of computers running software and human observers. One issue explored concerns the measurement process in quantum mechanics. Another issue explored concerns the nature of experience as revealed by the existence of dreams. Some difficulties stemming from a materialist philosophy as regards these issues are pointed out. For example, a gedankenexperiment involving what has been called "negative" observation is discussed that illustrates the difficulty with a materialist assumption in quantum mechanics. Based on an exploration of these difficulties, specifications are outlined briefly that would provide a means to demonstrate the equivalence of of computers running software and human experience given a materialist assumption.

## Text

There are those scholars, such as Dyer (1994), who maintain that there is no essential difference between a thinking person and a suitably programmed computer, as regards their cognitive capability. This belief is generally grounded in a materialist conception of existence. If both a person and a computer are ultimately describable in terms of the same type of physical interactions, then there is no reason for assuming a distinction between a thinking person and a suitably programmed computer as regards cognitive functioning. Even in quantum mechanics, where human cognitive involvement in the measurement process of physical phenomena is likely (due to the probabilistic nature of quantum mechanics and the in general immediate change in the wave function throughout space upon measurement of a physical existent), there is no essential difference between a thinking person and a suitably programmed computer, once the assumption of materialism is made.

The fact, though, is that materialism is an unwarranted assumption, an add-on to what quantum mechanics indicates is occurring. In quantum mechanics, the physical world does not exist independently of the person who knows it. The materialist assumption regarding the fundamental equivalence of human cognition and computer functioning due to their common material nature does not account for the intrinsic relationship between the physical world and the person who knows it in quantum mechanics. In a straightforward approach to quantum mechanics, the thinking, observing person is not reducible to the elements of the physical world because the person is part of a fundamental relation in which this person is one pole and the physical world is another pole in a basic relationship between *knower* and *known*. Quantum mechanics provides probabilistic predictions, in others words knowledge, without positing a deterministic world behind those predictions.

The situation is comparable in an important way to what Descartes maintained is the first indisputable fact, namely that an individual is aware of himself as a thinking being. In Descartes's system, the primacy of this first fact cannot reasonably be made contingent upon some other fact that is not more fundamental, for example the existence of the physical world. An essential difference between Descartes and a straightforward reading of quantum mechanics is that in quantum mechanics the observing, thinking individual and the physical world are *both* fundamental poles that exist in relation to one another.[1]

Quantum mechanics does not assume an independently existing physical world beyond what it allows an individual to know about the physical world. What it allows an individual to know are probabilities that measurements will yield certain outcomes. These probabilities themselves are dependent on previously made measurements. Dyer (1994) maintained that locality is not violated in quantum mechanics, which today generally means that causality is limited by the velocity of light in vacuum. From this position, one can argue "that fixed, objective properties can be assigned to quantum-level elements" (p. 285).

Contrary to Dyer's view, locality is violated in measurement in quantum mechanics. Rohrlich's (1983) paper in *Science* entitled, "Facing Quantum Mechanical Reality," begins with the statement, "Local hidden variables theory is dead" (p. 1251). The reason that faster-than-light communication is not possible is because of the statistical character of the physical events, but the violation of locality is manifest in comparing a number of spacelike separated events. When they are compared, correlations appear that do indicate a

---

[1] Descartes's position represents the opposite of materialism in that the primary fact is mind and the existence of the physical world is a deduced consequence.

# How Does a Computer

violation of locality.

The circumstances of measurement in quantum mechanics are unlike the generally held distinction between psychological and physical phenomena where the subjective aspects of perceptions of the physical world are ascribed to psychological phenomena independent of the physical world itself.  In measurement in quantum mechanics, the subjective aspect of perception concerns the physical world itself.  In view of the nature of measurement in quantum mechanics, one should inquire about the validity of a materialist assumption, including the corollary that computers and human beings are equivalent because each is a cognitive system possessing a material nature.

How do computers accomplish what human observers accomplish in making measurements in quantum mechanics?  Stated in Dyer's terms noted above, computers rely on fixed, objective properties in their basic operations, no matter how much flexibility is built into their operation at higher levels.  This is a key feature of materialism; the lack of such properties allows a fundamental role for subjectivity.  In contrast, measurement in quantum mechanics involves the lack of fixed, objective properties at the most basic level of functioning of the physical world.  The human observer is central to the nature of the physical world in quantum mechanics because: 1) it is a human observer making an observation who is linked to probabilistic predictions of quantum mechanics being realized in the physical world, and 2) the human observer is the agent who knows the probabilistic predictions that empirical evidence from the physical world indicates are correct.  How does a computer act as an observer when fundamentally the computer itself is a physical object and thus the object of the human observer's cognitive activity?  That is, even though on a macroscopic level, as has been noted, computers rely on fixed, objective properties in their basic operations, the most fundamental description of any physical system is in terms of quantum mechanical principles.  Dyer's adoption of materialism in his attempt to understand quantum mechanics is not supported by quantum mechanics itself.  The attempt to argue for the equivalence of human cognitive activity and the operation of a computer through a materialist philosophy is not supported in quantum mechanics.

A specific example involving what Renninger (1960) called "'negative' observation" (p. 418) in measurement in quantum mechanics will be presented shortly to support my contention that a materialist assumption concerning the observer in quantum mechanics is unwarranted and that consequently an argument like Dyer's concerning the equivalence a human observer and a computer running a program in quantum mechanical measurement is incorrect. First, though, a materialist assumption can be questioned without recourse to quantum mechanics by considering a central issue raised by Descartes regarding the nature of experience.

### DESCARTES'S DREAMS AND PROFESSOR DYER'S MATERIALISM

Descartes did not have an easy time with dreams in his attempt to establish the existence of the physical world. His work is instructive, this time with regard to the difficulty that those who maintain a materialist approach in general confront.

In reviewing his previous considerations concerning the dubitability of sensory experience, Descartes (1641/1969) wrote:

> But afterwards many experiences little by little destroyed all the faith which I had rested in my senses....And to those grounds of doubt I have lately added two others, which are very general; the first is that I never have believed myself to feel anything in waking moments which I cannot also sometimes believe myself to feel when I sleep, and as I do not think that these things which I seem to feel in sleep, proceed from objects outside of me, I do not see any reason why I should have this belief regarding objects which I seem to perceive while awake. The other was that being still ignorant, or rather supposing myself to be ignorant, of the author of my being, I saw nothing to prevent me from having been so constituted by nature that I might be deceived even in matters which seemed to me to be most certain....But since G-d is no deceiver, it is very manifest that He does not communicate to me these ideas immediately and by Himself [since this is not what I experience], nor yet by the intervention of some creature in which their reality is not formally, but only eminently, contained. For since He has given me no faculty to recognise that this is the case, but, on the other hand, a very great inclination to believe...that they are conveyed to me by corporeal objects, I do not see how He could be defended from the accusation of deceit if these ideas were produced by causes other than corporeal objects. Hence we must allow that corporeal objects exist. (pp. 212-213, 215)[2]

A materialist would not go through the same rational process as Descartes did. He might argue that his approach to experience is fundamentally different than that of Descartes, Descartes being a rationalist. It is true that the materialist's basic assumption regarding experience is that it can be understood in terms of one world and this world is material in nature. But, like the question for Descartes, which is the real world, or more precisely the material world? Is it the world in which a person is "awake" and in which he remembers the world of his dreams, even though he remembers that this "dream" world seemed just as "real" as the one in which he is "awake"? Furthermore, as Descartes said, "as I do not think [when I am awake] that these things which I seem to feel in sleep, proceed from objects outside of me, I do not see any reason why I should have this belief regarding objects which I seem to perceive while awake" (p. 213).

The point to be emphasized here is not with demonstrating that the physical world exists given the independence of mind from the physical world and the ontological priority of the mind, but rather how does a materialist know that his assumption of the "awake" world as the material world, the real world, is any more "real" than the world of dreams? More specifically, Dyer's computers are in the "awake, real" world, the material world. If this is the case, then how do these computers account for the human ability to dream, specifically to have experiences whose basic experiential validity is no different than the material world to which a materialist ascribes ontological priority?

Essentially, by adopting Descartes's conclusion that there is but one world and adding to it that everything that exists is fundamentally material in nature, Dyer can then assign elements in his computer program to elements in the physical world on a one-to-one basis. He can even attempt to account for subjectivity by assigning elements of his computer program to elements of the material (or physical) world on a many-to-one or one-to-many basis. As regards subjectivity, Dyer assumed some particular dynamics on the part of the cognitive entity that are reflected in the relation of the cognitive entity to the physical world. But Dyer did not account for the different *realms* of experience that dreaming indicates exist.

How does a materialist account for essentially different worlds

---
[2] The *o* in *G-d* has been removed.

# How Does a Computer

experienced by an observer, both with the same experiential validity, including one where computers need not exist? Different experienced worlds necessarily include the human observer engaged in experiencing them. This tenet, broad in its range of applicability, is particularly relevant to quantum mechanics where the thinking, perceiving person *as a subject* is fundamentally related to the physical world and cannot be reduced to the object perceived. In quantum mechanics, an observer's awareness concerning some physical existent is not a secondary phenomenon dependent on more fundamental materialistic processes, processes of the physical world independent of human cognition. (It is important to emphasize that the wave function associated with a physical existent generally changes immediately throughout space when the existent is measured. This change is not limited by the velocity of light in vacuum, the limiting velocity for physical existents in the special theory.) An individual's cognition, and his awareness in general, are related to the physical world in a fundamental manner. Where Descartes relied on the first fact of a person's awareness of his existence apart from the world, quantum mechanics indicates that the first fact is the observer's awareness of his inescapable relation to the world.

A materialist can always assume a person's perception, and the person's cognition in general, are dependent on some more fundamental material process and therefore argue that there is no fundamental difference between computers and human observers. To do so, though, raises significant problems, both conceptually and in terms of empirical evidence. In quantum mechanics, at the present time the materialist assumption does not have empirical support as concerns the equivalence of human and non-human observation. Furthermore, in his own struggle, Descartes argued for the existence of the physical world (really one physical world that yields only one truth), the world experienced when we are "awake," by relying on the existence of G-d. G-d, for Descartes "is no deceiver" (p. 215). I have no objection to Descartes's argument except that in science, empirical data are the determining factors in deciding the validity of a theory. Can Dyer provide empirical data that his assumption of materialism is correct? It is not enough to make the assumption and then develop computer programs to supposedly replicate what is occurring in human cognition. To support his assumption of materialism, Dyer must bring data that is independent of the human cognition that creates the programs (Snyder, 1985).

Here is an area where such an example would be particularly useful if Dyer, or another materialist, could develop one. It concerns what Renninger (1960) called a "'negative' observation." (p. 418). A "negative" observation is a change in the wave function associated with a physical existent that occurs when the time has elapsed for a physical existent to register a particular measurement result, and it has not occurred. The observer then knows that another result has occurred.

## FEYNMAN'S TWO-HOLE GEDANKENEXPERIMENTS

Generally the change in the wave function that often occurs in measurement in quantum mechanics has been ascribed to the unavoidable physical interaction between the measuring instrument and the physical entity measured. Indeed, Bohr (1935) maintained that this unavoidable interaction was responsible for the uncertainty principle, more specifically the inability to simultaneously measure observable quantities described by non-commuting Hermitian operators (e.g., the position and momentum of a particle). The following series of gedankenexperiments in this section will show that this interaction is not necessary to effect a change in the wave function. The series of gedankenexperiments indicates that knowledge plays a significant role in the change in the wave function that often occurs in measurement (Snyder, 1996a, 1996b).

*Gedankenexperiment 1*

Feynman, Leighton, and Sands (1965) explained that the distribution of electrons passing through a wall with two suitably arranged holes to a backstop where the positions of the electrons are detected exhibits interference (Figure 1). Electrons at the backstop may be detected with a Geiger counter or an electron multiplier. Feynman et al. explained that this interference is characteristic of wave phenomena and that the distribution of electrons at the backstop indicates that each of the electrons acts like a wave as it passes through the wall with two holes. It should be noted that when the electrons are detected in this gedankenexperiment, they are detected as discrete entities, a characteristic of particles, or in Feynman et al.'s terminology, "lumps" (p. 1-5).

In Figure 1, the absence of lines indicating possible paths for the electrons to take from the electron source to the backstop is not an oversight. An electron is not taking one or the other of the paths. Instead, the wave function associated with each electron after it passes through the holes is the

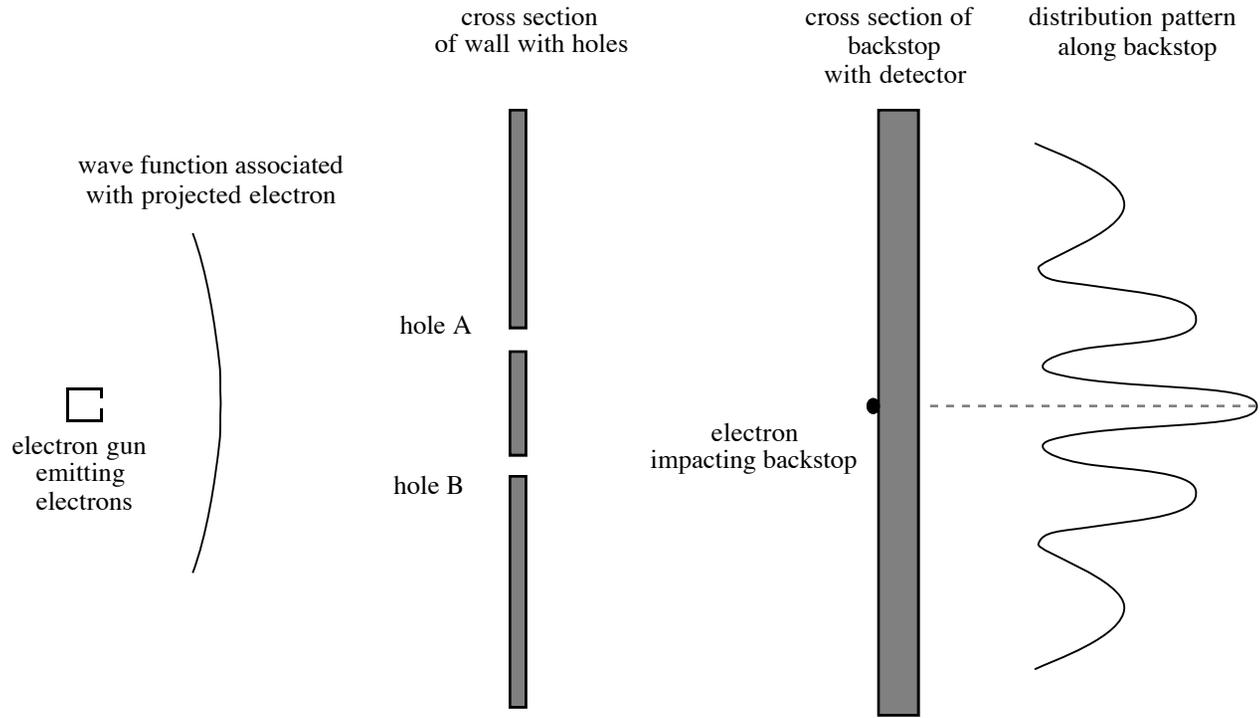

Figure 1

Two-hole gedankenexperiment in which the distribution of
electrons reflects interference in the wave functions of electrons.
(Gedankenexperiment 1)

sum of two more elementary wave functions, with each of these wave functions experiencing diffraction at one or the other of the holes. Epstein (1945) emphasized that when the quantum mechanical wave of some physical entity such as an electron exhibits interference, it is interference generated only in the wave function characterizing the individual entity.

The diffraction patterns resulting from the waves of the electrons passing through the two holes would at different spatial points along a backstop behind the hole exhibit constructive or destructive interference. At some points along the backstop, the waves from each hole sum (i.e., constructively interfere), and at other points along the backstop, the waves from each hole subtract (i.e., destructively interfere). The distribution of electrons at the backstop is given by the absolute square of the combined waves at different locations along the backstop, similar to the characteristic of a classical wave whose intensity at a particular location is proportional to the square of its amplitude. Because the electrons are detected as discrete entities, like particles, at the backstop, it takes many electrons to determine the intensity of the quantum wave that describes each of the electrons and that is reflected in the distribution of the electrons against the backstop.

*Gedankenexperiment 2*

Feynman et al. further explained that if one were to implement a procedure in which it could be determined through which hole the electron passed, the interference pattern is destroyed and the resulting distribution of the electrons resembles that of classical particles passing through the two holes in an important way. Feynman et al. relied on a strong light source behind the wall and between the two holes that illuminates an electron as it travels through either hole (Figure 2). Note the significant difference between the distribution patterns in Figures 1 and 2.

In Figure 2, the path from the electron's detection by the light to the backstop is indicated, but it is important to emphasize that this path is inferred only after the electron has reached the backstop. A measurement of the position of the electron with the use of the light source introduces an uncertainty in its momentum. Only when the electron is detected at the backstop can one infer the path the electron traveled from the hole it went through to the backstop. It is not something one can know before the electron strikes the backstop.

In Feynman et al.'s gedankenexperiment using the light source, the distribution of electrons passing through both holes would be similar to that

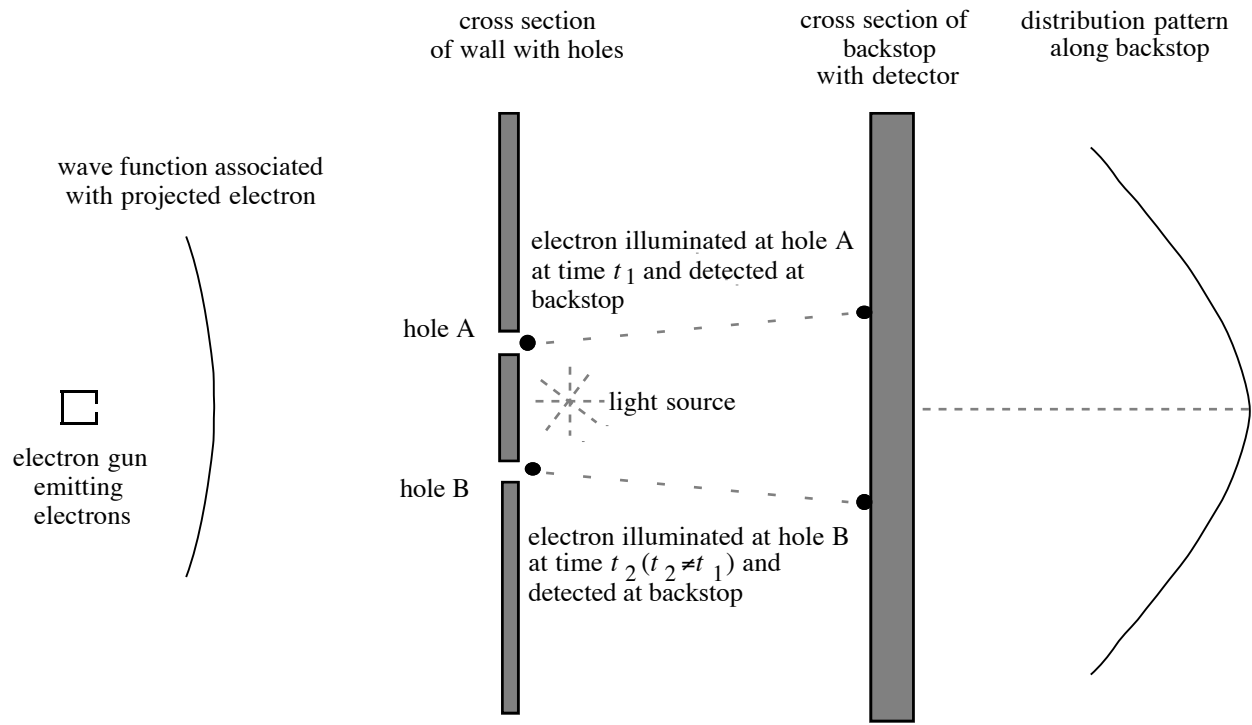

Figure 2
Two-hole gedankenexperiment with strong light source.
(Gedankenexperiment 2)

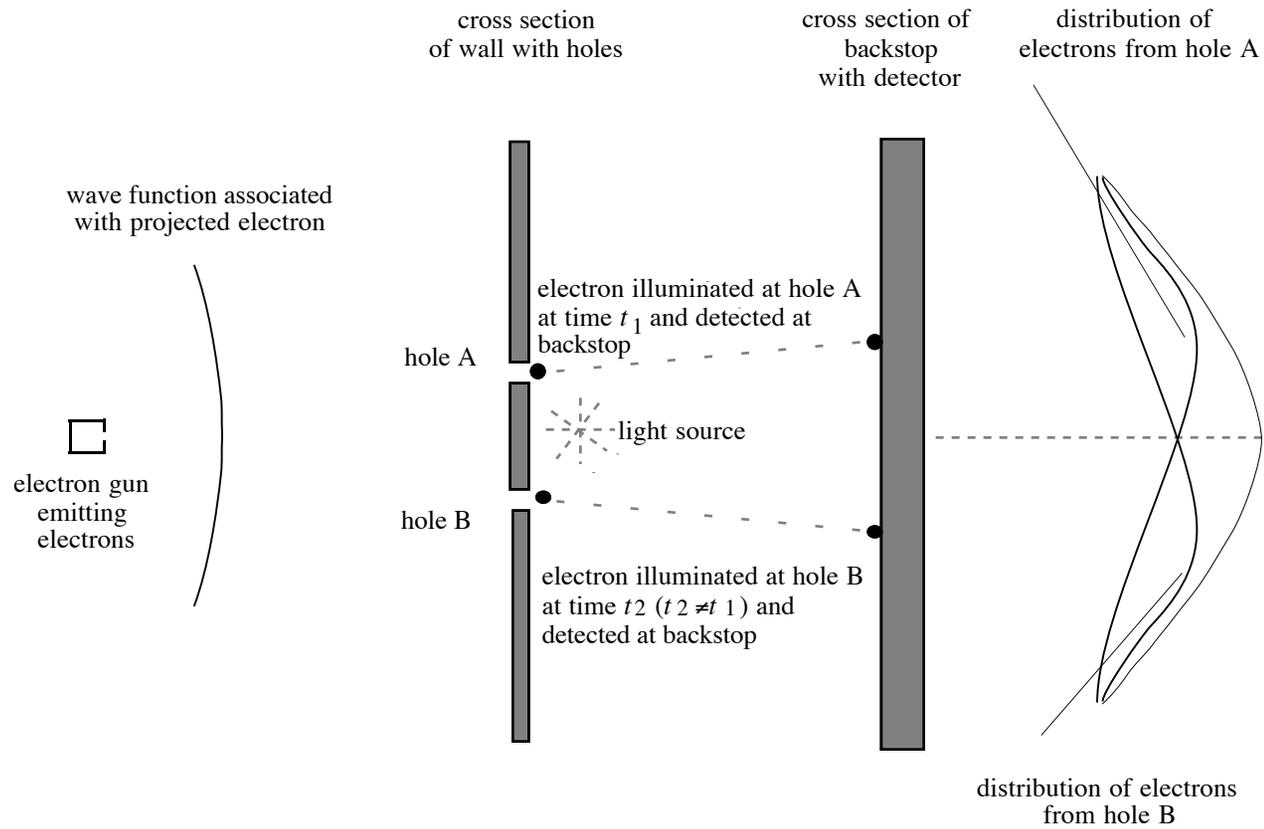

Figure 3

Two-hole gedankenexperiment with strong light source in which the distribution of electrons from each hole is shown.

found if classical particles were sent through an analogous experimental arrangement in an important way. Specifically, as in the case of classical particles, this distribution of electrons at the backstop is the simple summation of the distribution patterns for electrons passing through one or the other of the holes. Figure 3 shows the distribution patterns of electrons passing through hole A and electrons passing through hole B in Gedankenexperiment 2. These distribution patterns are identical to those that would occur if only one or the other of the holes were open at a particular time. An inspection of Figure 3 shows that summing the distribution patterns for the electrons passing through hole A and those passing through hole B results in the overall distribution of electrons found in Gedankenexperiment 2.

*The Uncertainty Principle*

Feynman et al.'s gedankenexperiments are themselves very interesting in that they illustrate certain apparently incongruent characteristics of microscopic physical existents, namely particle-like and wave-like features. Feynman et al. discussed their gedankenexperiments in terms of Heisenberg's uncertainty principle. Feynman et al. wrote:

> He [Heisenberg] proposed as a general principle, his *uncertainty principle*, which we can state in terms of our experiment as follows: "It is impossible to design an apparatus to determine which hole the electron passes through, that will not at the same time disturb the electrons enough to destroy the interference pattern." If an apparatus is capable of determining which hole the electron goes through, it *cannot* be so delicate that it does not disturb the pattern in an essential way. (p. 1-9)

Note that Feynman et al. implied in their description of the uncertainty principle that there is an unavoidable interaction between the measuring instrument (in their gedankenexperiment, the strong light source emitting photons) and the physical entity measured. Feynman et al. also wrote concerning Gedanken-experiment 2:

> the jolt given to the electron when the photon is scattered by it is such as to change the electron's motion enough so that if it might have gone to where $P_{12}$ [the electron distribution] was at a maximum [in Gedankenexperiment 1] it will instead land where $P_{12}$ was at a minimum; that is why we no longer see the wavy interference effects. (p. 1-8)

## How Does a Computer

In determining through which hole an electron passes, Feynman et al., like most physicists, maintained that the electrons are unavoidably disturbed by the photons from the light source and it is this disturbance by the photons that destroys the interference pattern. Indeed, in a survey of a number of the textbooks of quantum mechanics, it is interesting that each author, in line with Feynman and Bohr, allowed a central role in the change in the wave function that occurs in a measurement to a physical interaction between the physical existent measured and some physical measuring apparatus. The authors of these textbooks are Dicke and Witke (1960), Eisberg and Resnick (1974/1985), Gasiorowicz (1974), Goswami (1992), Liboff (1993), Merzbacher (1961/1970), and Messiah (1962/1965).

It is important to note explicitly that some causative factor is necessary to account for the very different distributions of the electrons in Figures 1 and 2. Feynman et al. maintained that the physical interaction between the electrons and photons from the light source is this factor.

*Gedankenexperiment 3*

Feynman et al.'s gedankenexperiments indicate that in quantum mechanics the act of taking a measurement in principle is linked to, and often affects, the physical world which is being measured. The nature of taking a measurement in quantum mechanics can be explored further by considering a certain variation of Feynman et al.'s second gedankenexperiment (Epstein, 1945; Renninger, 1960).[3] The results of this exploration are even more surprising than those presented by Feynman et al. in their gedankenexperiments. Empirical work on electron shelving that supports the next gedankenexperiment has been conducted by Nagourney, Sandberg, and Dehmelt (1986), Bergquist, Hulet, Itano, and Wineland (1986), and by Sauter, Neuhauser, Blatt, and Toschek (1986). This work has been summarized by Cook (1990).[4]

---

[3] Epstein (1945) presented the essence of Gedankenexperiment 3 using the passage of photons through an interferometer. Renninger (1960) also discussed a gedankenexperiment in an article entitled "Observations without Disturbing the Object" in which the essence of Gedankenexperiment 3 is presented.

[4] In electron shelving, an ion is placed into a superposition of two quantum states. In each of these states, an electron of the ion is in one or the other of two energy levels. The transition to one of the quantum states occurs very quickly and the transition to the other state occurs very slowly. If the ion is repeatedly placed in the superposition of states after it transitions to one or the other of the superposed states, one finds the atomic electron in general transitions very frequently between the superposed quantum states and the quantum state characterized by

In a similar arrangement to that found in Gedankenexperiment 2, one can determine which of the two holes an electron went through on its way to the backstop by using a light that is placed near only one of the holes and which illuminates only the hole it is placed by (Figure 4). Illuminating only one of the holes yields a distribution of the electrons similar to that which one would expect if the light were placed between the holes, as in Feynman et al.'s second gedankenexperiment. The distribution is similar to the sum of the distributions of electrons that one would expect if only one or the other of the holes were open at a particular time.

Moreover, when an observer knows that electrons have passed through the unilluminated hole because they were not seen to pass through the illuminated hole, the distribution of these electrons through the unilluminated hole resembles the distribution of electrons passing through the illuminated hole (Figure 5). Consider also the point that if: 1) the light is turned off before sufficient time has passed allowing the observer to conclude that an electron could not have passed through the illuminated hole, and 2) an electron has not been observed at the illuminated hole, the distribution of many such electrons passing through the wall is determined by an interference pattern that is the sum of diffraction patterns of the waves of the electrons passing through the two holes similar to that found in Gedankenexperiment 1 (Epstein, 1945; Renninger, 1960).

*Discussion of the Gedankenexperiments*

The immediate question is how are the results in Gedankenexperiment 3 possible given Feynman et al.'s thesis that physical interaction between the light source and electron is necessary to destroy the interference? Where the light illuminates only hole A, electrons passing through hole B do not interact with photons from the light source and yet interference is destroyed in the same manner as if the light source illuminated both holes A and B. In addition, the distribution of electrons passing through hole B at the backstop indicates that

---

the very quick transition. The photons emitted in these frequently occurring transitions to the quantum state characterized by the very quick transition are associated with resonance fluorescence of the ion. The absence of resonance fluorescence means that the ion has transitioned into the quantum state that occurs infrequently.

Cook (1990) has pointed out that in the work of Dehmelt and his colleagues on electron shelving involving the $Ba^+$ ion, the resonance fluorescence of a single ion is of sufficient intensity to be detectable by the dark-adapted eye alone, and the making of a negative observation, to be discussed shortly, is thus not dependent on any measuring device external to the observer.

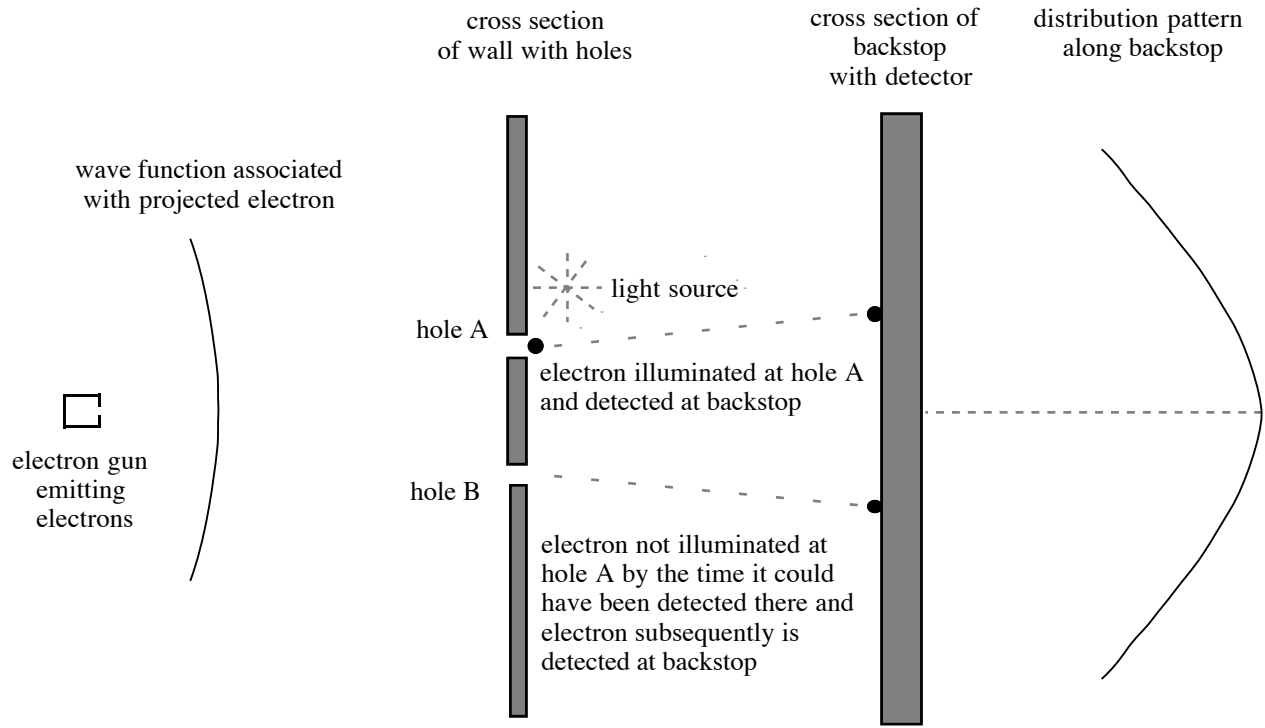

Figure 4

Two-hole gedankenexperiment with strong light source illuminating only one hole.
(Gedankenexperiment 3)

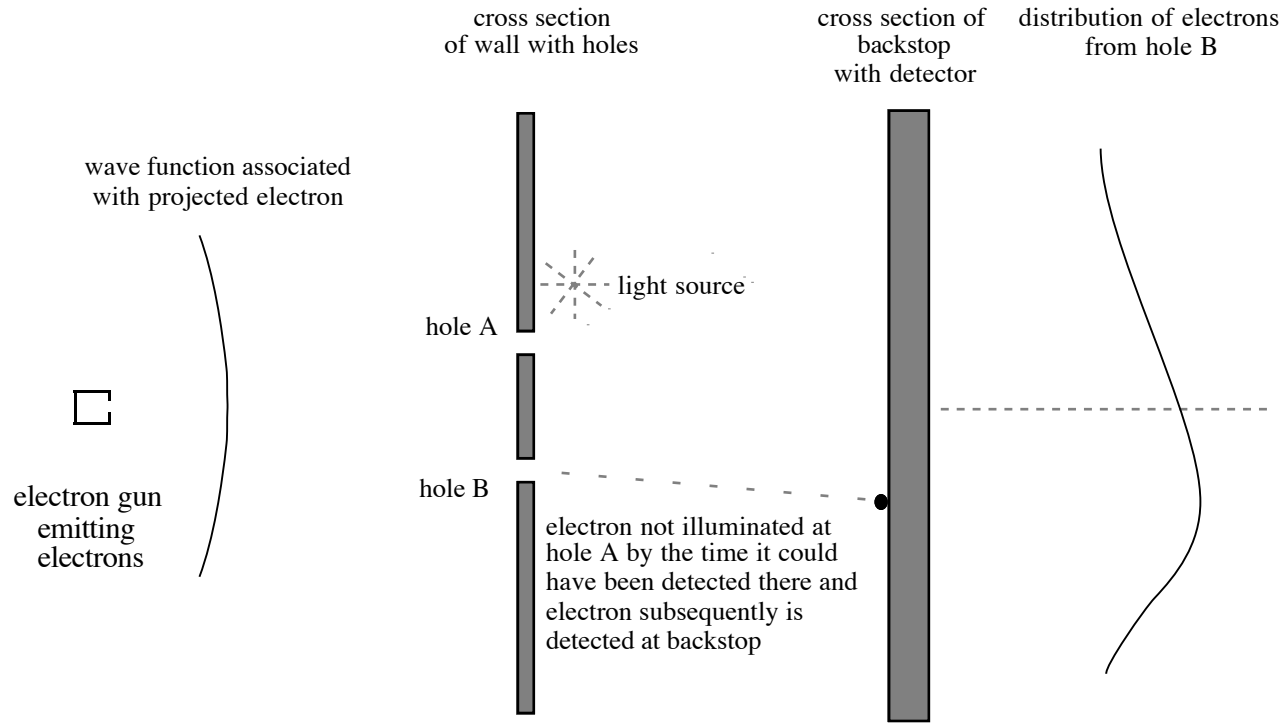

Figure 5

Two-hole gedankenexperiment with strong light source illuminating only one hole in which the distribution of electrons from unilluminated hole is shown.

there has been a change in the description of these electrons, even though no physical interaction has occurred between these electrons and photons from the light source.

Epstein (1945) maintained that these kinds of different effects on the physical world in quantum mechanics that cannot be ascribed to physical causes are associated with "*mental certainty*" (p. 134) on the part of an observer as to which of the possible alternatives for a physical existent occurs. Indeed, the factor responsible for the change in the wave function for an electron headed for holes A and B, and which is not illuminated at hole A, is *knowledge* by the observer as to whether there is sufficient time for an electron to pass through the "illuminated" hole. To borrow a term used by Renninger (1960), when the time has elapsed in which the electron could be illuminated at hole A, and it is not illuminated, the observer makes a "negative" (p. 418) observation.

The common factor associated with the electron's passage through the wall in a manner resembling that found for classical-like particles in Gedankenexperiments 2 and 3 is the observing, thinking individual's knowledge as to whether an electron passed through a particular hole. The physical interaction between photons from the light source and electrons passing through either hole 1 or hole 2 is not a common factor. It should be remembered that some causative factor is implied by the very different electron distributions in Gedankenexperiments 1 and 2. It is reasonable to conclude that knowledge by the observer regarding the particular path of the electron through the wall is a factor in the change in the distribution of the electrons in Gedankenexperiment 1 to that found for electrons in Gedankenexperiments 2 and 3.

It might be argued that in Gedankenexperiment 3 a non-human recording instrument might record whether or not an electron passed through the illuminated hole in the time allowed, apparently obviating the need for a human observer. But, as has been shown, a non-human recording instrument is not necessary to obtain the results in Gedankenexperiment 3. And yet even if a non-human instrument is used, ultimately a person is involved to read the results who could still be responsible for the obtained results. Furthermore, one would still have to explain the destruction of the interference affecting the distribution of the electrons at the backstop without relying on a physical interaction between the electrons and some other physical existent. Without ultimately relying on a human observer, this would be difficult to accomplish when the non-human recording instrument presumably relies on physical interactions for its functioning.

It should also be emphasized that the change in the wave function for an electron passing through the unilluminated hole in Gedankenexperiment 3 provides the general case concerning what is necessary for the change in a wave function to occur in a measurement of the physical existent with which it is associated. It was shown clearly in the extension of Feynman et al.'s gedankenexperiments that the change in the wave function of an electron or other physical existent is not due fundamentally to a physical cause. Instead, the change in the wave function is linked to the knowledge attained by the observer of the circumstances affecting the physical existent measured.

There is one other point to be emphasized. The change in the wave function discussed in Gedankenexperiment 3 serves only to capture the role of knowledge in negative observation. *That is, one need not even present a discussion of the wave function to attain the result that knowledge is a factor in the change in the electron distribution in Gedankenexperiment 1 to the electron distribution in Gedankenexperiments 2 and 3. This result depends only on the analysis of experimental results concerning the electron distributions in these three gedankenexperiments.*

CONCLUSION

Cognitive scientists, for example Dyer, provided evidence of the ability of computers running certain software to replicate certain features of the results of human cognition. The logical flow incorporated in the software can be considered by these scientists to characterize human thought itself. A computer program that replicates a "negative" observation in quantum mechanical measurement could be written, if one wished to do so. But what does one gain by running such a program on a computer? One does not gain insight into the nature of quantum mechanics, which can involve a human observer (more precisely an individual who can become aware of a certain set of circumstances) knowing that an event has not occurred during a certain period of time, with this knowledge tied to empirical results that differ from those that would result if he knew that this event had occurred. There is no physical interaction between the human observer and the physical existent measured, either direct in nature or reflected through the use of a macroscopic physical measuring instrument.

In a similar fashion, one could write a program that codes for two "worlds," an "awake" world, for example where computers exist, and a "dream" world, for example where computers do not exist. Again, what does one accomplish by doing this, except to represent features of human experience in the electric circuits of the computer? He does not add depth to our understanding of human experience. He does not account for essentially different worlds experienced by an observer, both having the same experiential validity.

Following Dyer's materialist assumption, as concerns measurement in quantum mechanics, if one wants to make a contribution in showing there is no fundamental difference between a computer running certain software and the human observer, he should specify how his computer running program, independent of a human observer, can destroy the interference in a wave function without physically interacting with the entity represented by the wave function and how people will know if the computer has accomplished this task. As concerns Descartes's dilemma regarding dreams, one could specify how a computer that is part of a material world can exist in a "dream" world in which a person can allow that computers do not even exist and which has the same experiential validity as the "awake" world or, in Dyer's terms, the material world.

## REFERENCES

Descartes, R. (1969). Meditations on first philosophy. In M D. Wilson (Ed.), *The essential Descartes*, pp. 154-223. New York: The New American Library. (Original work published 1641)

Dyer, M. G. (1994). Quantum physics and consciousness, creativity, computers: A commentary on Goswami's quantum-based theory of consciousness and free will. *The Journal of Mind and Behavior*, *15*, 265-290.

Epstein P. (1945). The reality problem in quantum mechanics. *American Journal of Physics*, *13*, 127-136.

Feynman, P. R., Leighton, R. B., &, and Sands, M. (1965). *The Feynman lectures on physics: Quantum mechanics* (Vol. 3). Reading, Massachusetts: Addison-Wesley.

Mermin, N. D. (1984). Spin correlation. *Science*, 223, 340-341.

Rohrlich, F. (1983). Facing quantum mechanical reality. *Science*, *221*, 1251-1255.

Renninger, M. (1960). Messungen ohne Störung des Meßobjekts [Observations without changing the object]. *Zeitschrift für Physik*, *158*, 417-421.

Snyder, D. M. (1985). On computer simulation of human cognition. *Psychological Reports*, *57*, 1317-1318.